\documentclass[a4paper]{mem}
\usepackage{txfonts}
\usepackage{natbib}
\usepackage{graphicx}
\usepackage[a4paper]{hyperref}
\idline{75}{1}
\begin{document}
   \title{Hard X-ray view of nearby Star Forming Regions}
   
   \author{S. Sciortino \inst{1,2}
}
   \offprints{S. Sciortino, sciorti@astropa.inaf.it}

   \institute{INAF-Osservatorio Astronomico di Palermo Giuseppe S. Vaiana,
Piazza del Parlamento 1, 90134 Palermo, Italy,  
 \email{sciorti@astropa.inaf.it} \\
     \and Consorzio COMETA, Via S. Sofia 64, Catania, Italy 
             }

\abstract{ {\em Chandra} and {\em XMM-Newton} have surveyed several nearby star forming regions and have greatly advanced our knowledge of X-ray emission from Young Stellar Objects (YSOs). After briefly reviewing it I discuss the advancements in this research field that could be possible with Simbol-X unique imaging capability in the hard X-ray bandpass.  
   }
   \authorrunning{S. Sciortino}
   \titlerunning{Hard X-ray view of nearby Star Forming Regions}
   \maketitle
%

\section{Introduction}

In the last decade thanks to the new space and ground 
instrumentations the study of star formation and early stellar evolution has gained a great impulse becoming a vibrant research field.

The role of X-rays, that likely trace, and are connected to, the magnetic
fields at work in the interaction region between the pre-main sequence star and its surrounding disk, has started to be recognized as one
of the relevant elements to deal with for understanding star 
formation and early evolution. Henceforth the X-ray observational window has gained an
increasing interest by the specialists in this research field. 

In the last few years the X-ray emission from Young Stellar Objects (YSOs) has been the subject of many {\em Chandra} and {\em XMM-Newton} studies that
have surveyed several nearby Star Forming Regions (SFRs) and have performed
deep observations of topical objects. Given the extent of the available 
literature it is impossible to review it 
in the space of this contribution, however most of our knowledge has
largely advantaged of  the results
of {\em COUP} ({\em C}handra {\em O}rion {\em U}ltradeep {\em P}roject,
cf. \citealt{ss_Getman+05}), of {\em XEST} ({\em X}MM-Newton {\em E}xtended
{\em S}urvey of {\em T}aurus, cf. \citealt{ss_Gudel+07}) and, more recently, of {\em DROXO} ({\em D}eep {\em R}ho {\em O}phiuchi {\em
X}MM-Newton {\em O}bservation, cf. \citealt{ss_Sciortino+06}).

Very synthetically our current knowledge on the YSO
X-ray emission can be summarized as follows: 
a) YSOs are X-ray luminous, at 1 M$_{sun}$ their typical X-ray emission is in the 
$\log(L_X~[{\rm erg/sec}])$ = 30$-$31 range. For comparison the typical value for the Sun is
$\log(L_X~[{\rm erg/sec}])$ = 26.5$-$27.5.
b) The emitting plasma is very hot. Plasma at several 10$^7$ K is always 
present and its temperature increases up to 10$^8$ K during flares.
For comparison, in the Sun plasma with temperature up to 10$^7$ K is present 
only during flares.
c) The emission is highly variable and, at the same time, can be present
stochastic variability, big flares (with intensity variations up to 
100 times), as well as rotational modulation.
d) The typical X-ray luminosity of Class~I/II YSOs is 2$-$3 times lower than that 
of Class~III YSOs.
e) Class~II YSO X-ray spectra show emitting plasma often with 
low metal abundances.
f) Class~I/II YSO have hotter X-ray spectra than Class III YSOs, 
but a soft (0.2-0.3 keV) component can be present. This latter component is associated
to accretion shocks due to disk material "falling" on the surface of the central accreting PMS star.

Seen in the perspective offered by Simbol-X, with its unique capacity to obtain, 
for the first time, real images of the X-ray sky
in the 10$-$60 keV bandpass, our current knowledge naturally brings to identify key open issues in
YSO studies, such as: i) the still controversial X-ray emission 
in Class~0 YSOs, i.e. in the early protostars with a lifetime of just
$\sim$ 10$^4$ yr;
ii) the existence of quiescent hard X-ray emission from Class~I-III YSOs
and its possible explanation in terms of continuous micro-flares; iii) the
occurrence of non-thermal hard X-ray emission during intense flares in YSOs and its possible analogy to the case of solar flares. All the 
above are connected to the general 
theme of the role played by high-energy radiation on star formation and early evolution. Other related issues are the channeling and regulation of accretion onto YSOs and the effects
of the high-energy radiation on the chemistry and early evolution of proto-planetary disks, as well as on the formation of complex molecules.

\begin{figure}
 \centering

 \includegraphics[clip=true,width=6cm]{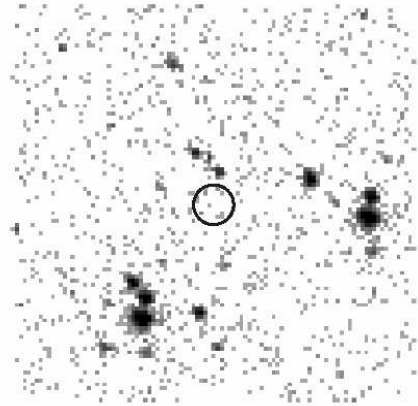}
 \caption{The staked image obtained by adding ACIS events in the 0.5-8.0 keV range from six regions of
200 $\times$ 200 pixels centered on the position of the 6 known Class~0 sources in the surveyed region.
The circular region in the center is 5", corresponding to positional
uncertainties of the mm/submm sources. 
No X-ray excess is present within this area indicating that the surveyed
Serpens Class~0 YSOs are unlikely
to be X-ray sources with intensities just below the detection threshold
(adapted from \citealt{ss_Giardinoa+07}).}
\label{fig6_ss}
\end{figure}

\section{X-rays from Class 0 YSOs}

\begin{figure}
 \centering
 \includegraphics[width=6cm]{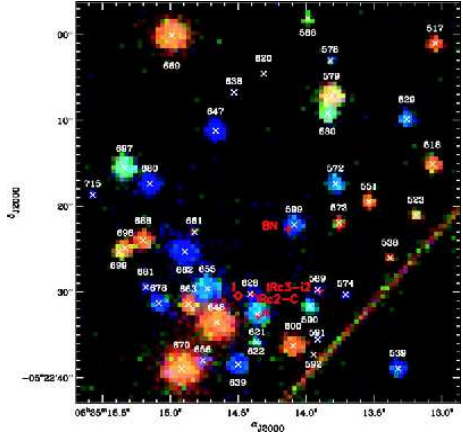} 
 \caption{A COUP color coded (red = 0.5-1.7 keV, green = 1.7-2.8, 
blue= 2.8-8.0 keV photons) view of the BN-KL region. Red crosses mark the 
positions of luminous mid-infrared sources. The red diamonds shows the radio 
position of source~I with no X-ray counterpart. Note the large population
of deeply embedded (blue) COUP sources (adapted from \cite{ss_Grosso+05}).}
\label{figCOUP_BN_ss}
\end{figure}

\begin{figure}[h]
 \centering
 \includegraphics[width=6cm]{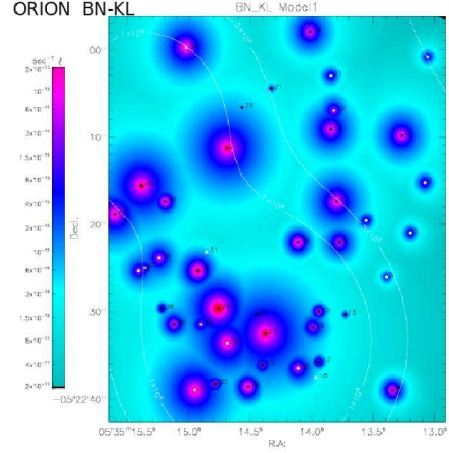}
 \caption{A two dimensional projection of a model computation of the
ionization rate as function of position for the BN cloud core 
in the OMC~1 region.  Across the entire core the (color coded)
value of ionization rate is higher than the typical value due to cosmic 
rays (2~10$^{-17}$ s$^{-1}$). Around each of the 
embedded X-ray emitting YSOs develops a R\"ontgen sphere where the 
X-ray induced
ionization rate is several orders of magnitude higher than the background level.
(courtesy of A. Lorenzani and F. Palla.)}
\label{fig6b_ss}
\end{figure}

After many years of search, the X-ray emission from Class~0 YSOs is still controversial. Either it is weak or rare or it is extremely difficult to discover due to
the conspicuous amount of intervening absorbing material.
While X-rays are quite
penetrating -- indeed the absorption at 2 keV and at 2 $\mu$m are similar -- Class~0 sources can be subject to extinction up to
100 magnitudes (and even higher) preventing the escape of any X-rays.  
One of the most (if not the most) stringent upper limit to the intrinsic 
X-ray luminosity of Class~0 has been obtained thanks
to a 100 ksec {\it Chandra} observation toward the
Serpens SFR \citep{ss_Giardinoa+07}. By staking data taken at 6 known Class~0 
positions it has been possible to determine that the Serpens 
Class~0 intrinsic X-ray luminosity is lower than 4~10$^{29}$ erg/s 
(assuming emission from an optically thin isothermal plasma with 
kT = 2.3 keV seen
through an absorbing column with N$_H$ = 4~10$^{23}$ cm$^{-2}$). The X-ray luminosity implyed by this upper limit is still
a dex higher than the X-ray luminosity of the active
Sun. Future deep observations with Simbol-X will make it possible to take
advantage, for
the first time, of the penetrating power of hard X-ray to really advance 
our knowledge on this subject. I feel this important because 
with COUP we have discovered in Orion a deeply embedded population
(cf. Fig. \ref{figCOUP_BN_ss}), that has been
shown to locally dominate the ionization level within a molecular cloud
(cf. \citet{ss_Lorenzani+07} and Fig. \ref{fig6b_ss}). However, as of
today, we do not know when intense X-ray emission from
YSOs develops and starts to affect -- for example by
determining the effectiveness of ambipolar diffusion -- the further evolution of star formation process. As a result  
we do not know yet if this effect is
just a small adjustment of the current interpretational 
scenario(s) or a major change is required if X-rays start acting at very early (Class~0) times.

\section{Flares and Magnetic Funnels}

The use of dynamical information (decay time, etc.) of flares is a classic tool to derive physical parameters, including size, of emitting regions (cf. Reale 2007). 
This is possible because in order to have a flare with the typical decay phase the plasma 
{\bf must} be confined (\cite{ss_Reale+02}). As a results the behavior of flare X-ray light curve (and the related time resolved spectra) allows measuring the size of flaring magnetic structure.
In normal stars the observed flares are similar to solar ones,
but sometimes much stronger (up to 10$^4$ times) both in absolute term and
with respect to the stellar bolometric luminosity. In most cases the observed YSO flares are similar to "standard" stellar ones, but there are a few notable exceptions: in about 10 {\em COUP} (Favata et al. 2005) and 2 {\em DROXO} (Flaccomio et al 2007) flares, the deduced size of the flaring region is at least 3 times larger than stellar radius and in few cases as long as 0.1 AU, i.e. the size of the star-disk separation. This long structures have never been "seen" in more evolved normal stars. Such long structures, if anchored on the star surface, will suffer severe stability problem
due to the centrifugal force since 1-2 Myr YOSs are fast rotators (P$_{rot}$ = 1-8 days) with a disk corotation radius of about 1-10 stellar radius. A possible solution, compatible with avaliable observational evidence, is one in which the loop connects the star with the disk at the corotation radius. Such magnetic funnels have been predicted by magnetospheric accretion model
(e.g., \citealt{ss_Shu+97}) and have been shown to occur in 
up-to-date MHD simulations of disk-star system (e.g., \citealt{ss_Long+07}), but it is only thanks to the {\em COUP} and {\em DROXO} long continuous observations that we have gained some observational evidence of their existence.

Returning for a moment to the "standard" flares, the detection of non-thermal hard X-rays during intense flares is discussed in this volume by 
\cite{ss_Maggio07} and \cite{ss_Argiroffi+07}, to which I defer the interested readers.

\section{The origin of the emission of Fe 6.4 keV K$_{\alpha}$ fluorescent line in YSOs}

\begin{figure}
 \centering
 \includegraphics[width=6cm]{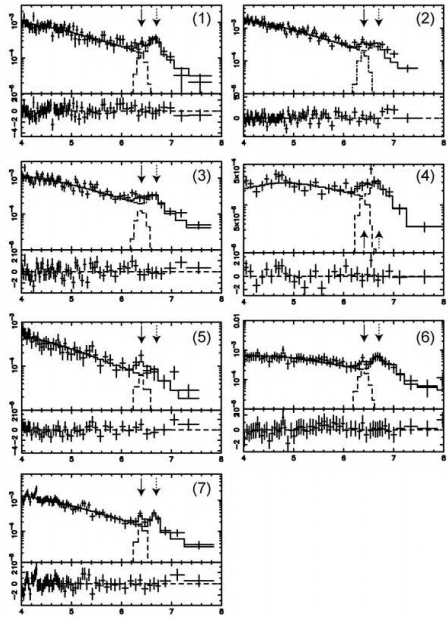}
 \caption{(Upper panels) Observed spectra (pluses) and best-fit models
(solid steps) of the 7 COUP YSOs showing the Fe fluorescent 6.4 keV
line. The Fe 6.4 keV line Gaussian component is shown by dashed steps.
The 6.4 and 6.7 keV lines are indicated by solid and broken arrows,
respectively. Photon energy in keV is on the abscissa, while
the ordinate is the spectral intensity as counts s$^{-1}$
keV$^{-1}$. (Lower panels) Residual to the fit in unit of $\chi$ values
(adapted from \citealt{ss_tu+05}).}
\label{fig7_ss} 
\end{figure}

The first detection of a Fe 6.4 keV K$_{\alpha}$ fluorescent line emission in a YSO
has been obtained with {\em Chandra} during an intense flare on YLW16A, a Class~I YSO 
in the $\rho$ Oph SFR \citep{ss_im+01}.
Thanks to COUP
we have collected 134 Orion YSOs spectra of sufficient
quality to allow investigating the spectra in
vicinity of the Fe XXV 6.7 keV line looking for the presence of the
Fe 6.4 keV K$_{\alpha}$ line.  In 7 COUP sources the 6.4 keV line
(cf. Fig \ref{fig7_ss}) has been found \citep{ss_tu+05} and the emission
has been interpreted, following original suggestion of \citet{ss_im+01},
as originating from the circumstellar neutral disk matter illuminated by
the X-ray emitted from the PMS star during the intense flares observed in all seven cases. A Fe 6.4 keV fluorescent line has
also been seen during a relatively short XMM-Newton observation of the
Class~II YSO Elias 29 without any evidence of concurrent flare emission
\citep{ss_Favata+05b}.
None or very limited time resolved spectroscopy
has been possible due either to the {\em XMM-Newton} too short observation or
the {\em Chandra} limited collecting area. Recently \citealt{ss_Czesla+Schmitt07}
have reported time-resolved spectroscopy of V1489 Ori, one of the 7 COUP 
sources with the Fe 6.4 keV K$_{\alpha}$ line, showing that this line appears 
predominantly in the 20 ks rise phase of a flare. Their preliminary calculation 
suggests that the photo-ionization alone cannot account for the observed
intensity of the Fe K$_{\alpha}$ line.

\begin{figure*}
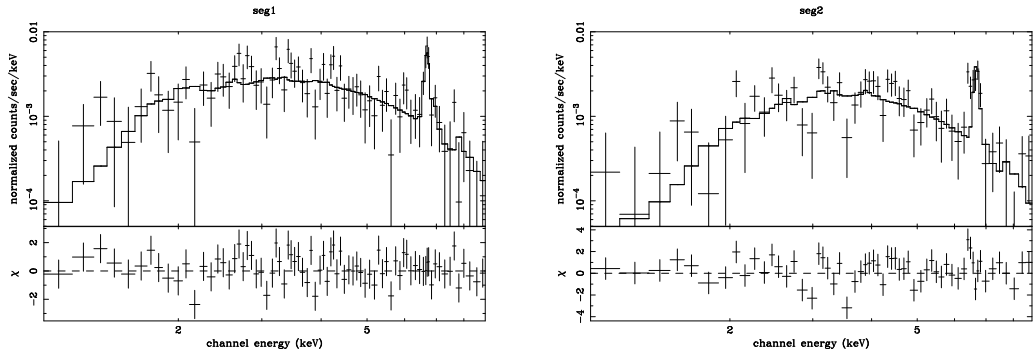

 \centering
 \resizebox{\hsize}{!}
 {{\rotatebox[]{-90}{\includegraphics[width=6cm,clip=true]{SSc_El29_seg1.ps}}}
\hspace{1cm}
 {\rotatebox[]{-90}{\includegraphics[width=6cm,clip=true]{SSc_El29_seg2.ps}}}}
 \vspace{-1.2cm}
 \caption{Spectra and spectral fit to the {\em DROXO} data of Elias 29 before the
flare (left) and after the flare (right). The spectra are
very similar in overall shape, intensity, and resulting best
fit model parameters.  After the flare,
however, a significant excess of emission at 6.4 keV is
present which is not visible in the data before the flare (adapted from
\citealt{ss_Giardinob+07}).}
\label{fig8_ss} 
\end{figure*}

Thanks to
{\em DROXO}, a XMM-Newton large program aiming to
study the properties of X-ray emission of the 1 Myr old $\rho$
Oph YSOs, it has been possible to perform, for the first
time, a time-resolved study of the Fe 6.4 keV fluorescent line emission
of Elias 29 \citep{ss_Giardinob+07}.  The line intensity is highly
variable. It is absent at the beginning of observation, then after a quite
typical flare
(a factor 8 in intensity with a 6 ksec decay time) it appears
with a conspicuous 
equivalent width, EW $\sim$ 250 eV. Subsequentely it continues to be present
with EW $\sim$ 150 eV
for the remaining 300 ksec (i.e. for 4 days!) of the observation. Apart for
the flare, the, relatively soft, X-ray spectra of Elias 29 remains essentially unchanged across
the entire observation, with no obvious hardening of the spectrum during 
the last 300 ksec of observation.
This behavior clearly challenges the "standard" interpretation of the
fluorescent emission as due to photo-ionizing X-ray photons (requiring
an adequate flux of photons with E $>$ 7.1 keV) and suggests an
alternative scenario in which the line is collisionally excited
by beams of electrons due to reconnections of magnetic field lines
occurring in the accretion funnels connecting the accreting pre-main-sequence star with its 
circumstellar disk. The required energy can be released by magnetic fields stressed near the 
corotation radius as a result of the radial gradient of rotational speed.

A deep Symbol-X observation will offer a unique opportunity to test 
this alternative scenario
since the presence of non-thermal electrons should be detectable through their 
bremsstrahlung radiation in the hard (Simbol-X) X-ray bandpass (while being 
invisible in the XMM observation). This is shown by the detailed simulations presented in this same volume by \cite{ss_Micela+07}.

\section{The quiescent hard X-ray emission of YSOs}

Another issue that could be explored with Simbol-X is the quiescent hard X-ray emission of YSOs and its origin and nature.

\begin{figure}
 \centering
 \includegraphics[width=6cm]{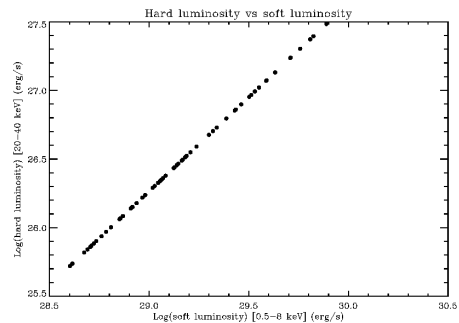}
 \caption{Predicted Hard (20-40 keV) vs. Soft (0.2-8.0 keV) X-ray luminosities for the putative YSOs similar to those found in ONC but at the typical distance ($\sim$ 150 pc) of a nearby star forming region. The
predicted hard X-ray emission allows the detection of those YSOs with Simbol-X.}
\label{fig_YSO_LX_ss}
\end{figure}

To set the framework let me remember that i) the analysis of the COUP light-curves of the Orion Nebula Cluster (ONC) low mass YSOs
has shown that their entire X-ray emission can be explained as due to continuous
flaring with flare intensities following a power law distribution characterized by a unique power law index 
\citep{ss_Caramazza+07} and ii) a recent analysis of both solar 
(RHESSI archive data) and few stellar flares has allowed to derive a
scaling law relation between soft thermal and hard non-thermal X-ray
emission \citep{ss_Isola+07a}. This relation holds over several dexs 
(see the figures in \cite{ss_Isola+07b} in this volume).

Combining these two empirical evidences we can predict the expected non-thermal hard X-ray flux as a function of the soft X-ray flux for YSOs like those studied in ONC by \cite{ss_Caramazza+07}: with a long Simbol-X observation (300-500 ksec) it would be possible to detect the hard X-ray emission due to the source continuous flaring. Assuming that similar sources will be hosted in nearby, $\sim$ 150 pc away, SFRs we can predict their expected hard X-ray luminosity (cf. Fig. \ref{fig_YSO_LX_ss}). Since the vast majority of YSOs have soft X-ray luminosity in the 10$^{29}$-10$^{30}$ erg/s range (or even higher) we expect that these putative sources would be detectable with the foreseen Simbol-X performances.

\section{Conclusions}

Long look observations of few selected SFRs with Simbol-X will offer the unique
opportunity to find whether Class~0 YSOs, ie. the very young 
(10$^4$ yr, still accreting) protostars, actually emit X-rays. Their emission, 
if present at a level comparable to those of older Class~I-III YSOs, could
have a significant impact on star formation and early evolution,
and may be also on the earliest stages of planetary formation
(e.g. large grains, planetesimals). These observations will also allow us to study the origin of the quiescent (if any) hard X-ray emission of YSOs. Finally, Symbol-X will open the possibility 
to study the MHD acceleration process in X-ray luminous YSOs that 
will offer a new, nearby,
laboratory for the study of other magnetic/shock related processes.

\begin{acknowledgements}
DROXO is a XMM-Newton Large Program (PI: S. Sciortino) 
supported by ASI-INAF contract I/023/05/0.
S. Sciortino acknowledges enlightening discussions
with many colleagues and their kind
share of results and figures in advance of final publication.
\end{acknowledgements}

\bibliographystyle{aa}

\begin{thebibliography}{}
\expandafter\ifx\csname natexlab\endcsname\relax\def\natexlab#1{#1}\fi

\bibitem[{Argiroffi {et~al.}(2007)}] {ss_Argiroffi+07}
Argiroffi, C., Micela, G., Maggio, A. 2007, Mem. Sait, in this volume

\bibitem[{Caramazza {et~al.}(2007)}] {ss_Caramazza+07} 
Caramazza, M., Flaccomio, E., Micela, G., Reale, F., Wolk, S. J., Feigelson, E. D. 2007, A\&A, 471, 645

\bibitem[{Czesla \& Schimtt(2007)}] {ss_Czesla+Schmitt07}
Czesla, S., Schmitt, J. H. H. M. 2007, A\&A, 470, L13

\bibitem[{Favata {et~al.}(2005a)}]{ss_Favata+05a} 
Favata, F., Flaccomio, E., Reale, F., Micela, G., Sciortino, S., et al., 2005a, ApJS, 160, 469

\bibitem[{Favata {et~al.}(2005b)}] {ss_Favata+05b}
Favata, F., Micela, G., Silva, B., Sciortino, S., Tsujimoto, M. 2005b, A\&A, 433, 104

\bibitem[{Flaccomio {et~al.}(2007)}]{ss_Flaccomio+07} 
Flaccomio, E., Stelzer, B., Pillitteri, I., Micela, G., Reale, F., Sciortino, S. 2007,
in Proc. of the "Cool Stars 14-th", (L. Rebull \& J. Stauffer, eds.), in press.

\bibitem[{Giardino {et~al.}(2007a)}]{ss_Giardinoa+07} 
Giardino, G., Favata, F., Micela, G., Sciortino, S., Wiston, E. 
2007a, A\&A, 463, 275

\bibitem[{Giardino {et~al.}(2007b)}]{ss_Giardinob+07} 
Giardino, G., Favata, F., Pillitteri, I., Flaccomio, E., Micela, G., \& 
Sciortino, S. 2007b, A\&A, in press

\bibitem[{Grosso {et al.}(2005)}] {ss_Grosso+05}
Grosso, N., Feigelson, E. D., Getman, K. V., et al. 2005, ApJS, 160, 530

\bibitem[{Getman {et~al.}(2005)}] {ss_Getman+05} 
Getman, K.~V., Flaccomio, E., Broos, P. S., et~al. 2005, ApJS, 160, 319

\bibitem[{G\"uedel {et~al.}(2007)}] {ss_Gudel+07}
G\"uedel, M.,Briggs, K. R., Arzner, K., et al. 2007, A\&A, 468, 353

\bibitem[{Imanishi {et~al.}(2001)}] {ss_im+01}
Imanishi, K., Koyama, K., Tsuboi, Y. 2001, ApJ, 557, 747

\bibitem[{Isola {et~al.}(2007a)}] {ss_Isola+07a}
Isola, C., Favata, F., Micela, G., Hudson, H. 2007, A\&A, 472, 261 

\bibitem[{Isola {et~al.}(2007b)}] {ss_Isola+07b}
Isola, C., Favata, F., Micela, G., Hudson, H. 2007, Mem. Sait, in this volume 

\bibitem[{Long {et~al.}(2007)}] {ss_Long+07}
Long, M., Romanova, M.~M., \& Lovelace, R.~V.~E. 2007, MNRAS, 374, 436

\bibitem[{Lorenzani {et~al.}(2007)}] {ss_Lorenzani+07}
Lorenzani, A., Palla, F., Feigelson, E.D., Grosso, N. 2007, in preparation

\bibitem[{Maggio(2007)}] {ss_Maggio07}
Maggio, A. 2007, Mem.Sait, in this volume

\bibitem[{Micela {et~al.}(2007)}] {ss_Micela+07}
Micela, G., Favata, F., Giardino, G., Sciortino, S. 2007, Mem.Sait, in this volume

\bibitem[{Reale(2007)}] {ss_Reale07}
Reale, F. 2007, A\&A, 471, 271 

\bibitem[{{Reale, Bocchino \& Peres}(2002)}] {ss_Reale+02}
Reale, F., Bocchino, F., Peres, G. 2002, A\&A, 383, 952

\bibitem[{Sciortino {et~al.}(2006)}] {ss_Sciortino+06}
Sciortino, S., Pillitteri, I., Damiani, F., et al. 2006, in 
Proc. of the "The X-ray Universe 2005", A. Wilson, (ed), ESA SP-604, 111 

\bibitem[{Shu {et~al.}(1997)}] {ss_Shu+97}
Shu, F.~H., Shang, H., Glassgold, A.~E., Lee, T. 1997, Science, 277, 1475

\bibitem[{Tusijmoto {et~al.}(2005)}] {ss_tu+05}
Tsuijmoto, M., Feigelson, E.~D., Grosso, N., et al. 2005, ApJS, 160, 503

\end{thebibliography}

\end{document}